\PassOptionsToPackage{unicode}{hyperref}
\PassOptionsToPackage{hyphens}{url}
\PassOptionsToPackage{dvipsnames,svgnames,x11names}{xcolor}
\documentclass[
]{article}
\usepackage{amsmath,amssymb}
\usepackage{lmodern}
\usepackage{iftex}
\ifPDFTeX
  \usepackage[T1]{fontenc}
  \usepackage[utf8]{inputenc}
  \usepackage{textcomp} 
\else 
  \usepackage{unicode-math}
  \defaultfontfeatures{Scale=MatchLowercase}
  \defaultfontfeatures[\rmfamily]{Ligatures=TeX,Scale=1}
\fi
\IfFileExists{upquote.sty}{\usepackage{upquote}}{}
\IfFileExists{microtype.sty}{
  \usepackage[]{microtype}
  \UseMicrotypeSet[protrusion]{basicmath} 
}{}
\makeatletter
\@ifundefined{KOMAClassName}{
  \IfFileExists{parskip.sty}{%
    \usepackage{parskip}
  }{
    \setlength{\parindent}{0pt}
    \setlength{\parskip}{6pt plus 2pt minus 1pt}}
}{
  \KOMAoptions{parskip=half}}
\makeatother
\usepackage{xcolor}
\usepackage{graphicx}
\makeatletter
\def\maxwidth{\ifdim\Gin@nat@width>\linewidth\linewidth\else\Gin@nat@width\fi}
\def\maxheight{\ifdim\Gin@nat@height>\textheight\textheight\else\Gin@nat@height\fi}
\makeatother
\setkeys{Gin}{width=\maxwidth,height=\maxheight,keepaspectratio}
\makeatletter
\def\fps@figure{htbp}
\makeatother
\NewDocumentCommand\citeproctext{}{}
\NewDocumentCommand\citeproc{mm}{%
  \begingroup\def\citeproctext{#2}\cite{#1}\endgroup}
\makeatletter
 \let\@cite@ofmt\@firstofone
 \def\@biblabel#1{}
 \def\@cite#1#2{{#1\if@tempswa , #2\fi}}
\makeatother
\newlength{\cslhangindent}
\newlength{\csllabelwidth}
\newenvironment{CSLReferences}[2] 
 {\begin{list}{}{%
  \setlength{\itemindent}{0pt}
  \setlength{\leftmargin}{0pt}
  \setlength{\parsep}{0pt}
  \ifodd #1
   \setlength{\leftmargin}{\cslhangindent}
   \setlength{\itemindent}{-1\cslhangindent}
  \fi
  \setlength{\itemsep}{#2\baselineskip}}}
 {\end{list}}
\usepackage{calc}

\ifLuaTeX
\usepackage[bidi=basic]{babel}
\else
\usepackage[bidi=default]{babel}
\fi
\babelprovide[main,import]{american}

\def\languageshorthands#1{}
\ifLuaTeX
  \usepackage{selnolig}  
\fi
\IfFileExists{bookmark.sty}{\usepackage{bookmark}}{\usepackage{hyperref}}
\IfFileExists{xurl.sty}{\usepackage{xurl}}{} 
\hypersetup{
  pdftitle={StationarityToolkit: Comprehensive Time Series Stationarity
Analysis in Python},
  pdfauthor={Bhanu Suraj Malla, Yuqing Hu},
  pdflang={en-US},
  colorlinks=true,
  linkcolor={Maroon},
  filecolor={Maroon},
  citecolor={Blue},
  urlcolor={Blue},
  pdfcreator={LaTeX via pandoc}}

\title{StationarityToolkit: Comprehensive Time Series Stationarity
Analysis in Python}

\definecolor{c53baa1}{RGB}{83,186,161}
\definecolor{c202826}{RGB}{32,40,38}

\usepackage[affil-it]{authblk}
\usepackage{orcidlink}
\author[1%
  \ensuremath\mathparagraph]{Bhanu Suraj Malla%
    \,\orcidlink{0009-0000-3593-2455}\,%
    }
\author[1%
  ]{Yuqing Hu%
    \,\orcidlink{0009-0008-3071-473X}\,%
    }

\affil[1]{Georgia Institute of Technology, United States%
  }
\affil[$\mathparagraph$]{Corresponding author: %
}
\date{5 March 2026}

\begin{document}
\maketitle

\section{Summary}\label{summary}

Time-series stationarity is a property that statistical characteristics
such as trend, variance, seasonality remain constant over time. It is
considered fundamental to many forecasting and analysis methods.
Different tests detect different types of non-stationarity: structural
breaks or deterministic trends, clustered or time-dependent variance,
stochastic or deterministic seasonality. A series might pass one test
while failing another; single-test approaches seldom distinguish between
conceptually different types of non-stationarity that require different
types of tests and transformations.

\texttt{StationarityToolkit} addresses this by providing a comprehensive
Python library that runs 10 statistical tests across three categories:
trend (4 tests), variance (4 tests), and seasonality (2 tests). Rather
than a binary stationary/non-stationary verdict, users receive detailed
diagnostics with actionable notes for each detection. The toolkit
automatically infers the frequency of the data provided (requires
datetime index), provides clear interpretations with test statistics and
p-values, and supports an iterative test-transform-retest workflow
essential for real-world data sets.

\section{Statement of Need}\label{statement-of-need}

Stationarity testing is one of the critical preprocessing steps for time
series analysis, but it is not a single question with a single answer.
Non-stationarity can manifest as unit roots, deterministic trends,
structural breaks, variance instability, volatility clustering, or
seasonal patterns - each requiring different transformations.
Practitioners need to identify \emph{which type} of non-stationarity
exists to choose the right remedy: differencing for unit roots
(\citeproc{ref-dickey1979unit}{Dickey \& Fuller, 1979}), detrending for
deterministic trends, Box-Cox (\citeproc{ref-box1964transformations}{Box
\& Cox, 1964}) for variance instability, or seasonal differencing for
periodic patterns.

This gap becomes particularly challenging when transformations interact
unpredictably. Differencing to remove trend can introduce variance
non-stationarity; variance stabilization can be undone by subsequent
differencing. Without comprehensive testing after each transformation,
practitioners cannot verify whether their preprocessing actually
achieved stationarity or introduced new problems. An iterative
test-transform-retest workflow is essential, but orchestrating this
across multiple libraries and tests is tedious and error-prone.

Potential users include data scientists, econometricians, and
researchers working with time series data who need to prepare data for
different use cases including ARIMA/SARIMA models, VAR analysis, Granger
causality tests, or machine learning applications where stationarity
improves generalization. The toolkit is particularly valuable for
practitioners who need to understand \emph{what type} of
non-stationarity exists in their data.

\section{State of the Field}\label{state-of-the-field}

Several Python packages address aspects of time series stationarity
testing. The \texttt{statsmodels} library
(\citeproc{ref-seabold2010statsmodels}{Seabold \& Perktold, 2010})
provides individual unit root tests (ADF
(\citeproc{ref-dickey1979unit}{Dickey \& Fuller, 1979}), KPSS
(\citeproc{ref-kwiatkowski1992kpss}{Kwiatkowski et al., 1992}),
Phillips-Perron (\citeproc{ref-phillips1988testing}{Phillips \& Perron,
1988}), Zivot-Andrews (\citeproc{ref-zivot1992further}{Zivot \& Andrews,
1992})) and seasonal decomposition (STL
(\citeproc{ref-cleveland1990stl}{Cleveland et al., 1990})). The
\texttt{arch} package (\citeproc{ref-sheppard2017arch}{Sheppard, 2017})
offers unit root tests and ARCH/GARCH models
(\citeproc{ref-engle1982arch}{Engle, 1982}) for volatility modeling,
though its primary focus is on fitting volatility models rather than
comprehensive stationarity diagnostics. The \texttt{scipy} library
(\citeproc{ref-2020SciPy-NMeth}{Virtanen et al., 2020}) offers variance
comparison tests (Levene (\citeproc{ref-levene1960robust}{Levene,
1960}), Bartlett (\citeproc{ref-bartlett1937properties}{Bartlett,
1937})). The \texttt{pmdarima} library
(\citeproc{ref-smith2017pmdarima}{Smith, 2015}) includes stationarity
tests primarily as preprocessing for auto-ARIMA model selection rather
than as standalone diagnostic tools. In all cases, users must manually
run tests from different libraries, interpret potentially conflicting
results, and determine which transformations to apply.

\texttt{StationarityToolkit} differs by integrating testing across all
the stationarity dimensions - trend, variance, and seasonality - in a
single call with a report that summarizes test outcome, actionable
notes, caveats, and their statistical interpretation. It goes one step
further to infer time-series frequency (requires datetime index) before
testing for seasonal non-stationarity to maintain intuitiveness in its
results. This approach also reveals test limitations in its notes (e.g.,
Zivot-Andrews detecting ``breaks'' in smooth trends, ARCH triggering on
auto-correlation) that single-test approaches miss.

The design philosophy of the toolkit prioritizes transparency over
prescription; the toolkit shows users what's happening in their data
rather than making transformation decisions for them. This is because
transformation effectiveness could vary unpredictably across datasets,
and across use cases as demonstrated in the package documentation where
identical transformations produce opposite variance outcomes on
synthetic versus real data.

\section{Software design}\label{software-design}

\texttt{StationarityToolkit} is implemented as a pure Python package
with dependencies on \texttt{numpy}
(\citeproc{ref-2020NumPy-Array}{Harris et al., 2020}), \texttt{pandas}
(\citeproc{ref-mckinney2010pandas}{McKinney, 2010}), \texttt{scipy}
(\citeproc{ref-2020SciPy-NMeth}{Virtanen et al., 2020}), \texttt{arch}
(\citeproc{ref-sheppard2017arch}{Sheppard, 2017}), and
\texttt{statsmodels} (\citeproc{ref-seabold2010statsmodels}{Seabold \&
Perktold, 2010}). The architecture separates test implementation (in
\texttt{tests/} modules organized by trend, variance, and seasonality),
result formatting and output generation (\texttt{results.py}), and the
main detection entry point (\texttt{toolkit.py}). This design enables
community contributions, reliably expanding the suite of tests while
maintaining consistent user experience.

The toolkit evolved significantly from its initial design. Early
versions (0.x) attempted to provide both testing and automated
transformations with inverse functions, but this approach proved
problematic as the effectiveness of transformations was found to vary
unpredictably across datasets, and their needs vary by use case; this
meant automated decisions obscured what was actually happening to the
data. Version 1.x and the following 2.x represented a fundamental
redesign, pivoting to a scaled-down pure diagnostic analysis tool with
an expanded test suite grounded in statistical theory. This design
choice - diagnostics over automation - was based on the reflection about
the reality that practitioners need to understand their data's specific
non-stationarity characteristics to make informed decisions for their
use case.

Key design decisions include:

\begin{enumerate}
\def\labelenumi{\arabic{enumi}.}
\item
  \textbf{Comprehensive testing by default}: All of the 10 statistical
  tests are run in a single call eliminating the cognitive burden on
  deciding which tests to run. This also ensures that the users don't
  miss critical information that they would otherwise miss.
\item
  \textbf{Structured output with actionable notes}: Each test returns
  not just pass/fail but also concise notes on the test, its caveats and
  suggestions (e.g.~``Unit root detected - consider differencing'',
  ``Deterministic trend detected - stationary after detrending'').
\item
  \textbf{Contextual seasonality detection}: Seasonal tests
  automatically determine appropriate periods to test based on time
  series frequency (e.g.~testing for weekly, monthly, and yearly
  seasonality in daily data), eliminating the need for manually
  specifying periods to test. This requires the time series to have
  datetime index.
\item
  \textbf{2-pronged reporting}: The \texttt{report()} method returns
  results as a pandas DataFrame and optionally writes a markdown report
  when a filepath is provided, supporting both interactive analysis and
  documentation.
\item
  \textbf{Intuitive trend tests}: All unit root tests (ADF
  (\citeproc{ref-dickey1979unit}{Dickey \& Fuller, 1979}), KPSS
  (\citeproc{ref-kwiatkowski1992kpss}{Kwiatkowski et al., 1992}),
  Phillips-Perron (\citeproc{ref-phillips1988testing}{Phillips \&
  Perron, 1988})) run to test both constant-only and constant-plus-trend
  parameters, intuitively determining whether non-stationarity is due to
  unit roots or deterministic trends.
\end{enumerate}

The implementation prioritizes correctness and interpretability. Tests
run sequentially and the typical execution time for all tests on a
1000-row series is under 2 seconds on modern hardware.

\begin{figure}
\centering
\includegraphics[width=1\linewidth,height=\textheight,keepaspectratio]{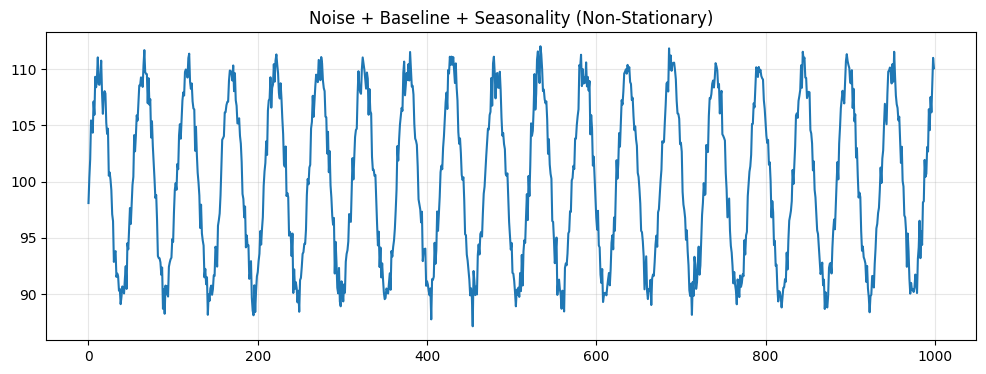}
\caption{Example input: synthetic time series with noise, baseline, and
seasonality.}
\end{figure}

\begin{figure}
\centering
\includegraphics[width=1\linewidth,height=\textheight,keepaspectratio]{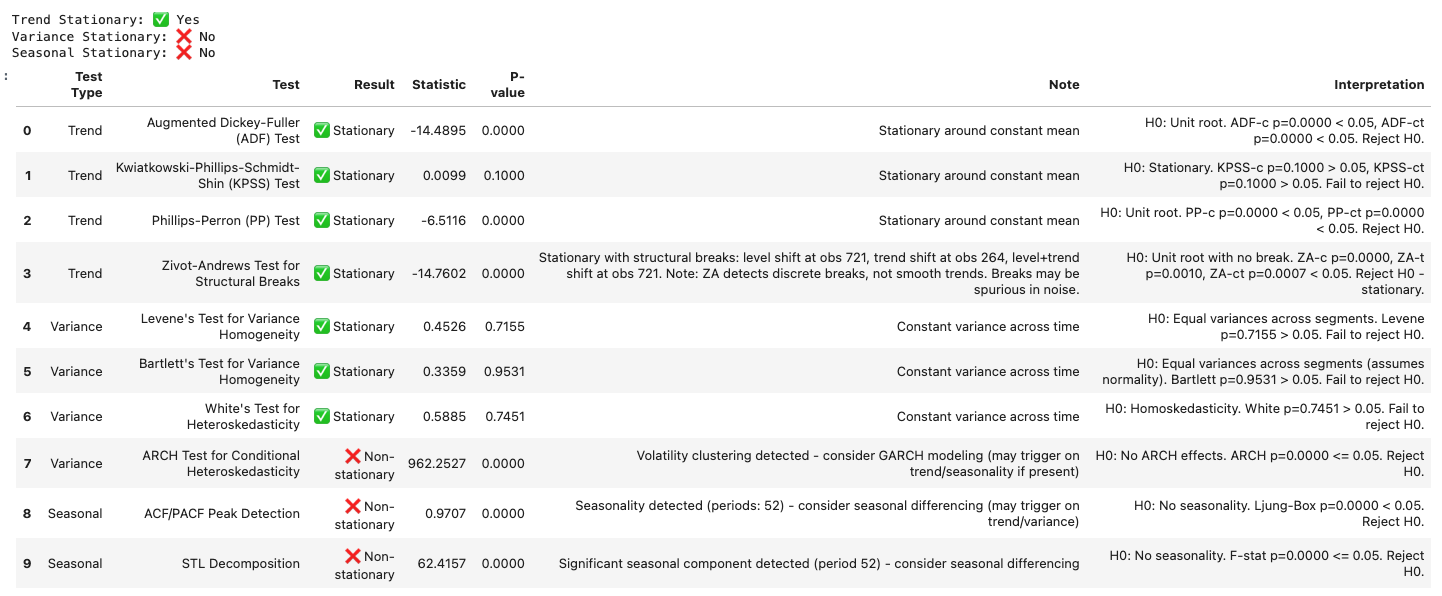}
\caption{Example output: StationarityToolkit summary and detailed test
results DataFrame.}
\end{figure}

\section{Research Impact Statement}\label{research-impact-statement}

\texttt{StationarityToolkit} addresses a genuine gap in the Python
ecosystem; while individual stationarity tests exist in various
libraries, no tool provides comprehensive integrated testing across
trend, variance, and seasonality dimensions with actionable notes and
caveats. The package has been publicly available on PyPI since 2023.

The toolkit's research significance lies in its ability to reveal test
limitations and cross-contamination effects (e.g.~variance
non-stationarity emerging as a result of differencing to allay trend
non-stationarity) that single-test approaches miss. The comprehensive
documentation, including validation on synthetic as well as real data
(\texttt{examples/detailed\_usage.ipynb}), provides researchers with a
context for when and why different tests succeed or fail. This
transparency is particularly valuable when reproducibility is important
- where understanding \emph{why} a transformation was chosen matters as
much as the transformation itself.

The package demonstrates credible significance through its complete
implementation of established statistical tests, clear documentation
with reproducible examples, and active maintenance. It is designed for
immediate integration into research workflows, with pandas DataFrame
outputs that integrate seamlessly with existing analysis pipelines and
markdown export for documentation.

\section{AI Usage Disclosure}\label{ai-usage-disclosure}

The toolkit's core design, tests write-up, examples and this paper were
all conceived and written by human author. The specific use of AI is as
follows:

\textbf{Tool use}: Claude (Anthropic, versions 3.5 Sonnet and Opus 4)
was used for refining code, docs.

\textbf{Nature and scope of assistance}: AI assisted with removing
redundant code (written by human), changing variable names (for
intuitiveness), adding doc-strings, and documentation (refining README
and notebooks inside \texttt{examples/}). Also, the idea of using
dataclass for \texttt{results.py} was assisted by AI with focus to
improve the result synthesis for users.

\textbf{Confirmation of review}: All AI-generated content was reviewed,
edited, and validated by the human author. Core design decisions were
made by the human author based on domain expertise in time series
analysis. Test implementations were validated against established
libraries (\texttt{statsmodels}, \texttt{arch}, \texttt{scipy}) and on
synthetic data with known properties. The human author takes full
responsibility for the accuracy, correctness, and scientific validity of
all submitted materials.

\section*{References}\label{references}
\addcontentsline{toc}{section}{References}

\protect\phantomsection\label{refs}


\begin{CSLReferences}{1}{0}
\bibitem[\citeproctext]{ref-bartlett1937properties}
Bartlett, M. S. (1937). Properties of sufficiency and statistical tests.
\emph{Proceedings of the Royal Society A}, \emph{160}(901), 268--282.

\bibitem[\citeproctext]{ref-box1964transformations}
Box, G. E. P., \& Cox, D. R. (1964). An analysis of transformations.
\emph{Journal of the Royal Statistical Society: Series B}, \emph{26}(2),
211--252.

\bibitem[\citeproctext]{ref-cleveland1990stl}
Cleveland, R. B., Cleveland, W. S., McRae, J. E., \& Terpenning, I.
(1990). STL: A seasonal-trend decomposition procedure based on loess.
\emph{Journal of Official Statistics}, \emph{6}(1), 3--33.

\bibitem[\citeproctext]{ref-dickey1979unit}
Dickey, D. A., \& Fuller, W. A. (1979). Distribution of the estimators
for autoregressive time series with a unit root. \emph{Journal of the
American Statistical Association}, \emph{74}(366a), 427--431.

\bibitem[\citeproctext]{ref-engle1982arch}
Engle, R. F. (1982). Autoregressive conditional heteroscedasticity with
estimates of the variance of united kingdom inflation.
\emph{Econometrica}, \emph{50}(4), 987--1007.

\bibitem[\citeproctext]{ref-2020NumPy-Array}
Harris, C. R., Millman, K. J., Walt, S. J. van der, \& others. (2020).
Array programming with NumPy. \emph{Nature}, \emph{585}(7825), 357--362.
\url{https://doi.org/10.1038/s41586-020-2649-2}

\bibitem[\citeproctext]{ref-kwiatkowski1992kpss}
Kwiatkowski, D., Phillips, P. C. B., Schmidt, P., \& Shin, Y. (1992).
Testing the null hypothesis of stationarity against the alternative of a
unit root. \emph{Journal of Econometrics}, \emph{54}(1-3), 159--178.

\bibitem[\citeproctext]{ref-levene1960robust}
Levene, H. (1960). Robust tests for equality of variances. In I. Olkin
(Ed.), \emph{Contributions to probability and statistics} (pp.
278--292). Stanford University Press.

\bibitem[\citeproctext]{ref-mckinney2010pandas}
McKinney, W. (2010). Data structures for statistical computing in
python. \emph{Proceedings of the 9th Python in Science Conference},
\emph{445}, 56--61. \url{https://doi.org/10.25080/Majora-92bf1922-00a}

\bibitem[\citeproctext]{ref-phillips1988testing}
Phillips, P. C. B., \& Perron, P. (1988). Testing for a unit root in
time series regression. \emph{Biometrika}, \emph{75}(2), 335--346.

\bibitem[\citeproctext]{ref-seabold2010statsmodels}
Seabold, S., \& Perktold, J. (2010). Statsmodels: Econometric and
statistical modeling with python. \emph{9th Python in Science
Conference}.

\bibitem[\citeproctext]{ref-sheppard2017arch}
Sheppard, K. (2017). \emph{Arch: ARCH models in python}. Zenodo.
\url{https://doi.org/10.5281/zenodo.593254}

\bibitem[\citeproctext]{ref-smith2017pmdarima}
Smith, T. G. (2015). \emph{Pmdarima} (Version 2.1.1).
\url{https://github.com/alkaline-ml/pmdarima}

\bibitem[\citeproctext]{ref-2020SciPy-NMeth}
Virtanen, P., Gommers, R., Oliphant, T. E., \& others. (2020). SciPy
1.0: Fundamental algorithms for scientific computing in python.
\emph{Nature Methods}, \emph{17}(3), 261--272.
\url{https://doi.org/10.1038/s41592-019-0686-2}

\bibitem[\citeproctext]{ref-zivot1992further}
Zivot, E., \& Andrews, D. W. K. (1992). Further evidence on the great
crash, the oil-price shock, and the unit-root hypothesis. \emph{Journal
of Business \& Economic Statistics}, \emph{10}(3), 251--270.

\end{CSLReferences}
\end{document}